\documentstyle[12pt,fleqn]{article}
\def\beq{\begin{equation}}
\def\eeq{\end{equation}}
\def\Eq{Eq.~(\ref}
\def\0{\otimes}
\def\1{\mbox{\small1\hskip-0.35em\normalsize1}_}
\def\2{\sqrt{2}}
\def\5{\sigma_}
\def\6{\langle}
\def\9{\rangle}
\def\Ph{{\mit\Phi}^}
\def\Ps{{\mit\Psi}^}

\begin{document}

\newcommand{\PRL}{{\sl Phys.\ Rev.\ Lett.\/}, }
\newcommand{\PRA}{{\sl Phys.\ Rev.\/} A, }

\vspace*{15mm}
\noindent{\Large {\bf Delayed choice for entanglement swapping}}

\vspace{1cm}\hspace*{14.5mm}\parbox{140mm}{ASHER PERES\\[5mm]
Department of Physics,\\ Technion---Israel Institute of
Technology,\\ 32 000 Haifa, Israel\\[2cm]
{\bf Abstract.} \ Two observers (Alice and Bob) independently prepare
two sets of singlets. They test one particle of each singlet along an
arbitrarily chosen direction and send the other particle to a third
observer, Eve. At a later time, Eve performs joint tests on pairs of
particles (one from Alice and one from Bob). According to Eve's choice
of test and to her results, Alice and Bob can sort into subsets the
samples that they have already tested, and they can verify that each
subset behaves as if it consisted of entangled pairs of distant
particles, that have never communicated in the past, even indirectly via
other particles.}\vfill

\noindent To appear in a special issue of \ {\sl Journal of Modern
Optics\/}.\vfill

\newpage\noindent{\bf1. \ Introduction and notations}

Since the early days of quantum mechanics, it has been known that after
two quantum systems interact, their joint state is usually entangled and
it may remain so even after these systems separate and are far away from
each other. However, a direct interaction is not necessary in order to
produce entanglement between distant systems. For example, any existing
entanglement between two particles can be teleported to other, distant
particles, by performing suitable joint measurements and broadcasting
their results as classical information~[1]. Protocols known as {\it
entanglement swapping\/} have recently been proposed [2,~3] and
experimentally realized~[4]. In the present article, I propose an even
more paradoxical experiment, where entanglement is produced {\it a
posteriori\/}, after the entangled particles have been measured and may
no longer exist.

To simplify the discourse and the notations, the experiment will be
described in terms of spin-$1\over2$ particles. (In the real world,
it would be easier to use polarized photons. The spin sphere then has to
be understood as a Poincar\'e sphere, and the argument is exactly the
same.) The eigenstates of $\5z$ will be denoted as

\beq {1\choose0}\equiv|0\9\qquad\qquad\mbox{and}\qquad\qquad
   {0\choose1}\equiv|1\9. \eeq
In that basis, the other spin components have the following nonvanishing
matrix elements:

\beq \60|\5x|1\9=\61|\5x|0\9=1, \qquad\qquad
 \60|\5y|1\9=-\61|\5y|0\9=-i. \label{xy}\eeq
We shall also need the spin components along directions at 45$^\circ$
from the $x$ and $y$ axes:

\beq \5\pm=(\5x\pm\5y)/\2. \eeq
These give

\beq \60|\5\pm|1\9=\61|\5\mp|0\9=(1\mp i)/\2. \label{uv}\eeq

For a pair of particles, it is convenient to define ``Bell states''~[5]

\beq \Ph\pm=(|00\9\pm|11\9)/\2\qquad\qquad\mbox{and}\qquad\qquad
 \Ps\pm=(|01\9\pm|10\9)/\2. \eeq
Note that $\Ps-$ is the singlet state.

In the proposed experiment, two distant observers, conventionally called
Alice and Bob, independently prepare two sets of singlets, whose states
are denoted as $\Ps-_A$ and $\Ps-_B$. Alice and Bob keep one particle
of each singlet and send the other particle to a third observer, Eve,
who also arranges them in pairs (one from Alice and one from Bob). The
three observers keep records specifying to which pair each particle
belongs.

The joint state of a pair of singlets can thus be written as

\beq \Ps-_A\0\Ps-_B\equiv
 (\Ps+_E\0\Ps+-\Ps-_E\0\Ps--\Ph+_E\0\Ph++\Ph-_E\0\Ph-)/2,
\label{swap}\eeq
where the subscript $E$ refers to the particles that are sent to Eve,
and the symbols $\Ph\pm$ and $\Ps\pm$ without a subscript refer to the
particles that Alice and Bob keep. Note the minus signs in \Eq{swap}).
It would be possible to eliminate them by using, instead of the Bell
basis, the ``magic basis'' [6,~7] where $\Ph-$ and $\Ps+$ are multiplied
by $i$.\bigskip

\noindent{\bf2. \ Analysis of the results of measurements}

Alice and Bob now measure the values of spin components (along arbitrary
directions) of the particles that they kept. For example, Alice measures
spin component $\5x$ or $\5y$ (randomly chosen) of her particles, and
likewise Bob measures $\5\pm$ on his particles. These components were
chosen because Alice and Bob want to test Bell inequalities, in the form
given by Clauser, Horne, Shimony, and Holt (CHSH)~[8]. The results that
Alice and Bob get, namely $\pm1$, are of course completely random and
uncorrelated.

At a later time, Eve performs joint tests on her pairs of particles.
Just as in the teleportation scenario~[1], she performs Bell
measurements~[5] and she informs the other observers of the results that
she found. Using that information, Alice and Bob sort the records of
their measurements into four subsets, according to whether Eve found
$\Ph\pm$ or $\Ps\pm$. It then follows from \Eq{swap}) that, in each
subset, the state of the particles that Alice and Bob kept was the same
as the state later found by Eve. Thus, in each subset, there are
nonvanishing expectation values $\6\Ph\pm|\5a\0\5b|\Ph\pm\9$ and
$\6\Ps\pm|\5a\0\5b|\Ps\pm\9$. Explicitly, owing to Eqs.~(\ref{xy}) and
(\ref{uv}), we have

\beq \6\Ps+|\5x\0\5\pm|\Ps+\9=-\6\Ps-|\5x\0\5\pm|\Ps-\9=1/\2, \eeq
\beq \6\Ps+|\5y\0\5\pm|\Ps+\9=-\6\Ps-|\5y\0\5\pm|\Ps-\9=\pm1/\2, \eeq
\beq \6\Ph+|\5x\0\5\pm|\Ph+\9=-\6\Ph-|\5x\0\5\pm|\Ph-\9=1/\2, \eeq
\beq \6\Ph+|\5y\0\5\pm|\Ph+\9=-\6\Ph-|\5y\0\5\pm|\Ph-\9=\mp1/\2. \eeq
Thus, in each subset, one of the following CHSH inequalities is violated
(and the others are satified):

\beq -2\leq\6\5x\0\5++\5x\0\5-+\5y\0\5+-\5y\0\5-\9\geq2, \label{psi}\eeq
\beq -2\leq\6\5x\0\5++\5x\0\5--\5y\0\5++\5y\0\5-\9\geq2. \label{phi}\eeq
For the subset associated with $\Ps-_E$, it is the left hand side of
\Eq{psi}) that is violated, for $\Ps+_E$, it is the right hand side. For
the subset associated with $\Ph-_E$, it is the left hand side of
\Eq{phi}) that is violated; and for $\Ph+_E$, it is the right hand side.
In other words, Alice and Bob find experimentally that each one of the
four postselected subsets produces statistical results identical to
those arising from maximally entangled pairs.\bigskip

\noindent{\bf3. \ The paradox}

There can be no doubt that the particles that were independently
produced and tested by Alice and Bob were uncorrelated and therefore
unentangled. Each one of these particles may well have disappeared
(e.g., been absorbed) before the next particle was produced, and before
Eve performed her tests. Only the records kept by the three observers
remain, to be examined objectively.

How can the appearance of entanglement arise in these circumstances? The
point is that it is meaningless to assert that two particles are
entangled without specifying in which state they are entangled, just as
it is meaningless to assert that a quantum system is in a pure state
without specifying that state~[9]. If this simple rule is forgotten, or
if we attempt to attribute an objective meaning to the quantum state of
a single system, curious paradoxes appear: quantum effects mimic not
only instantaneous action-at-a-distance but also, as seen here,
influence of future actions on past events, even after these events have
been irrevocably recorded.

Note in particular that even after Alice and Bob have recorded the
results of all their measurements, Eve still has the freedom of deciding
which experiment she will perform. It can be a Bell measurement as
proposed above, but the latter can also be preceded by arbitrary
bilateral rotations of the two spin-$1\over2$ particles (this
corresponds to an arbitrary real orthogonal transformation of the magic
basis [6,~7]). Eve can also perform an incomplete Bell measurement
[10--12], or any other measurement she decides, represented by any
positive operator valued measure (POVM)~[9] of her choice. The only
demand is that at least one of her outcomes corresponds to a definite
entangled state of the {\it other\/} pair of particles, namely those
retained by Alice and Bob (though not necessarily a maximally entangled
state, nor even a pure state of these particles)~[13].\medskip

\noindent{\bf Proof:} \ Let $\1E$ and $\1{AB}$ denote the unit matrices
(of order~4) for Eve's pairs and for the particles kept by Alice and
Bob. Let the elements of Eve's POVM be denoted as $E^\mu$ (satisfying
$\sum_\mu E^\mu=\1E$). Then the joint state of Alice and Bob's
particles, that corresponds to outcome $\mu$ registered by Eve, is the
partial trace

\beq \rho^\mu_{AB}=\mbox{Tr}_E\,[\rho\,(E^\mu\0\1{AB})], \eeq
where $\rho$ is the initial state, given by \Eq{swap}). Note that
$\rho^\mu_{AB}$ is not normalized: its trace is the probability that Eve
observes outcome $\mu$. The above relationship readily follows from the
fact that if $\{F^k\}$ any POVM chosen by Alice and Bob, with $\sum_k
F^k=\1{AB}$, then the probability of the joint result $\mu k$ is
$\mbox{Tr}\,[\rho\,(E^\mu\0F^k)]$.\medskip

It is not even necessary for Alice and Bob to know which experiments Eve
will do. If they know nothing, they just measure $\5x$, $\5y$, or $\5z$
(randomly chosen) on each one of their particles, instead of the
specific components that appear in Eqs.~(\ref{xy}) and (\ref{uv}).
Later, they will learn from Eve that a definite subset of her
experiments ascertained the existence of a definite entangled state of
their particles. Alice and Bob don't even have to know which state this
was. They will simply set apart the results of their prior measurements
for the corresponding subset of particles and compute the correlations
$\6\5a\0\5b\9$. From the latter, they can obtain the density matrix
$\rho$ of the pairs of particles in that subset~[14]. Subjecting that
density matrix to a partial transposition~[15], they will find that the
latter has a negative eigenvalue, thus verifying that the corresponding
subset of particles, if it still existed, would have an entangled state.

It is obvious that from the raw data collected by Alice and Bob it is
possible to select in many different ways subsets that correspond to
entangled pairs. The only role that Eve has in this experiment is to
tell Alice and Bob how to select such a subset. Clearly, Eve has to be
honest: if she does not perform her measurements in the correct way and
if she reports fake data, Alice and Bob will not select good subsets,
and then their analysis will readily expose Eve's misbehaviour.

In summary, there is nothing paradoxical in the experiments outlined
above. However, one has to clearly understand quantum mechanics and to
firmly believe in its correctness to see that there is no paradox.

\vspace{10mm}\noindent{\bf Acknowledgments}

I am grateful to Chris Fuchs for improving the presentation of this
article. This work was supported by the Gerard Swope Fund and the Fund
for Encouragement of Research.\newpage


\end{document}